\documentclass[aps,prb,reprint,superscriptaddress]{revtex4-2}

\usepackage{graphicx}
\usepackage{dcolumn}
\usepackage{bm}
\usepackage{amsmath}
\usepackage{amssymb}
\usepackage{graphicx}
\usepackage{indentfirst}
\usepackage{booktabs}
\usepackage{multirow}
\usepackage{colortbl}
\usepackage[T1]{fontenc}
\usepackage{textcomp}

\emergencystretch=\maxdimen
\begin{document}


\title{Unconventional iron-magnesium compounds at terapascal pressures}


\author{Yimei Fang}
\affiliation{Department of Physics, OSED,\\
Key Laboratory of Low Dimensional Condensed Matter Physics\\ (Department of Education of Fujian Province)\\
Jiujiang Research institute, Xiamen University, Xiamen 361005, China.}
\author{Yang Sun}
\email{ys3339@columbia.edu}
\affiliation{Department of Applied Physics and Applied Mathematics, Columbia University, New York, NY, 10027, USA}
\author{Renhai Wang}
\affiliation{Department of Physics, University of Science and Technology of China, Hefei 230026, China}
\affiliation{Department of Physics, Iowa State University, Ames, Iowa 50011, United States}
\author{Feng Zheng}
\affiliation{Department of Physics, OSED,\\
Key Laboratory of Low Dimensional Condensed Matter Physics\\ (Department of Education of Fujian Province)\\
Jiujiang Research institute, Xiamen University, Xiamen 361005, China.}
\author{Shunqing Wu}
\email{wsq@xmu.edu.cn}
\affiliation{Department of Physics, OSED,\\
Key Laboratory of Low Dimensional Condensed Matter Physics\\ (Department of Education of Fujian Province)\\
Jiujiang Research institute, Xiamen University, Xiamen 361005, China.}
\author{Cai-Zhuang Wang}
\affiliation{Department of Physics, Iowa State University, Ames, Iowa 50011, United States}
\author{Renata M. Wentzcovitch}
\email{rmw2150@columbia.edu}
\affiliation{Department of Applied Physics and Applied Mathematics, Columbia University, New York, NY, 10027, USA}
\affiliation{Department of Earth and Environmental Sciences, Columbia University, New York, NY, 10027, USA}
\affiliation{Lamont-Doherty Earth Observatory, Columbia University, Palisades, NY, 10964, USA}

\author{Kai-Ming Ho}
\affiliation{Department of Physics, Iowa State University, Ames, Iowa 50011, United States}

\date{\today}

\begin{abstract}
Being a lithophile element at ambient pressure, magnesium is long believed to be immiscible with iron. A recent study by Gao et al.~\cite{gao2019iron} showed that pressure turns magnesium into a siderophile element and can produce unconventional Fe-Mg compounds. Here, we extend the investigation to exoplanetary pressure conditions using an adaptive genetic algorithm-based variable-composition structural prediction approach. We identify several Fe-Mg phases up to 3 TPa. Our cluster alignment analysis reveals that most of the predicted Fe-Mg compounds prefer a BCC packing motif at terapascal pressures. This study provides a more comprehensive structure database to support future investigations of the high-pressure structural behavior of Fe-Mg and ternary, quaternary, etc. compounds involving these elements.
\end{abstract}


\maketitle

\section{Introduction}
For systems with significant atomic size mismatch at ambient conditions, limited solid inter-solubility is observed. One such system is the Fe-Mg binary alloy. Previous results showed that below 1273 K, Mg does not dissolve in Fe, while at the liquidus temperature, the maximum solubility of Mg in $\delta$-Fe only reaches 0.25 atomic percent (at.\%)~\cite{okamoto1990binary}. Some attempts have been made to facilitate Fe-Mg inter-alloying using ion-beam mixing~\cite{jaouen1989ion} or mechanical alloying~\cite{yelsukov2005mechanism}. Besides, several studies shave shown that high pressures can improve the Fe-Mg inter-solubility. At 20 GPa and 2273 K, Dubrovinskaia \emph{et al.} achieved a homogeneous Fe-Mg alloy with 4 at.\% Mg~\cite{dubrovinskaia2004iron}. Later on, the same authors observed a significantly improved solubility of Mg ($>$ 10 at.\%) in Fe at 126(3) GPa and 3650(250) K[6]. The authors ascribed the improved Fe-Mg inter-solubility to the dramatic atomic size difference reduction under pressure~\cite{dubrovinskaia2005beating}.

There are also various theoretical investigations on the possibility of Fe-Mg inter-alloying under Earth\textquotesingle s core conditions. Kadas et al. demonstrated that Mg plays an essential role in \emph{bcc} Fe\textquotesingle s dynamical stability and that a \emph{bcc} structured Fe-Mg alloy with 5-10 at.\% Mg reproduces the physical properties of Earth\textquotesingle s inner core very well~\cite{kadas2009stability}. Li \emph{et al.} found that solid Fe can incorporate substantial amounts of Mg at 360 GPa and 6500 K~\cite{li2018mg}. More recently, Gao \emph{et al.} predicted a series of stable Fe-Mg compounds with different stoichiometries under pressures up to 360 GPa [1]. An analysis of the electron localization function and density of states of these Fe-Mg compounds indicated that the electron transfer from Mg to Fe helps the formation of Fe-Mg compounds at high pressures~\cite{gao2019iron}. These theoretical findings suggest that Mg is a likely light element in the Earth\textquotesingle s solid core.

To date, limited studies have reported the formation of Fe-Mg compounds under exoplanetary interior pressures. Here, we perform an adaptive genetic algorithm (AGA) based structure prediction of the binary Fe-Mg phase diagram at 1TPa, 2TPa, and 3TPa. Several unexpected compounds, i.e., Fe$_2$Mg, FeMg, FeMg$_2$, and FeMg$_3$ are found to be stable. By exploring the local packing motifs of stable and metastable compounds, we find the BCC packing motif is favored at high pressure. Our current study focuses on the structural and motif information. Temperature effects on the stability of newly found phases will be addressed in a future study.

In the following section, we describe the computational details of structural prediction method and the density functional theory (DFT) calculations. Section III shows the identified new phases and their stability, as well as discussions of the results. Conclusions are presented in Sec. IV

~\clearpage
\section{Computational Methods}
The structural prediction of Fe-Mg compounds was carried out using an adaptive genetic algorithm (AGA) which offers a balance between the speed of structure exploration with classical potentials and the accuracy of DFT calculation in an iterative way. The initial candidate structure pool in the GA search was generated by randomly creating 128 structures without any assumption on the lattice symmetry. The structures were then relaxed to the nearest local minima and ranked by their enthalpies. In each GA generation, 32 new structures, i.e.,1/4 of the pool size, were produced from the parent structure pool through the mating procedure described in Ref.\cite{deaven1995molecular}. The new structures replaced the worst 32 structures in the pool to form a new generation of structures. We performed structure searches for 600 consecutive GA generations under each set of auxiliary interatomic potential. After the GA search cycle, 16 lowest-enthalpy structures were selected for DFT calculations to produce enthalpies, forces, and stresses for re-adjusting the classical auxiliary potential parameters for the next GA search. A total of 40 adaptive iterations were performed to obtain the final structures for the given chemical composition. Here, the classical auxiliary potential was determined by the embedded-atom method (EAM) \cite{foiles1986embedded} based interatomic potentials. Within EAM, the total energy of an N-atom system has the form
\begin{equation}
E_{total}=\frac{1}{2}\sum\nolimits_{i,j(i\ne j)}^{N}\phi(r_{ij})+\sum\nolimits_{i}F_{i}(n_i)
\end{equation}
where $\phi(r_{ij})$ denotes the pair repulsion between atoms $i$ and $j$ with a distance of $r_{ij}$, $F_{i}(n_i)$ is the embedded term with electron density term $n_i=\sum\nolimits_{j \ne i}\rho_{j}(r_{ij})$ at the site occupied by atom $i$. The fitting parameters in the EAM formula for the Fe-Mg system are determined as follows: the Lennard-Jones function modeled the parameters for Fe-Fe, Fe-Mg, and Mg-Mg interactions,
\begin{equation}
\phi(r_{ij}) = 4\varepsilon[(\frac{\sigma}{r_{ij}})^{12}-(\frac{\sigma}{r_{ij}})^{6}],
\end{equation}
where $\varepsilon$ and $\sigma$ are the fitting parameters. For Fe and Mg atoms, the density function was modeled by an exponentially decaying function
\begin{equation}
\rho(r_{ij}) = \alpha exp[-\beta(r_{ij}-r_{0})],
\end{equation}
$\alpha$ and $\beta$ are fitting parameters, and the embedding function takes the form proposed by Benerjea and Smith in Ref.\cite{banerjea1988origins} as follows:
\begin{equation}
F(n) = F_0[1-\gamma {\rm ln} n]n^{\gamma},
\end{equation}
where F$_0$ and $\gamma$ are fitting parameters. During the AGA run, the fitting parameters were adjusted adaptively in the light of the DFT calculated enthalpies, forces, and stresses of selected structures. The fitting procedure was realized using the force-matching method with the stochastic simulated annealing algorithm implemented in the POTFIT code~\cite{brommer2006effective,brommer2007potfit}. The first-principles calculations were carried out utilizing the Quantum ESPRESSO (QE) code \cite{giannozzi2009quantum,giannozzi2017advanced}. The exchange-correlation functional was treated with the non-spin-polarized generalized-gradient approximation (GGA) and parameterized by the Perdew-Burke-Ernzerhof formula (PBE). A kinetic-energy cutoff of 50 Ry for wave functions and 500 Ry for potentials were used. Brillouin-zone integration was performed over k-point grid of 2$\pi\times$ 0.03~\AA$^{-1}$ in the structure refinement. The convergence thresholds are 0.01 eV/\AA~ for the atomic force, 0.5 kbar for the pressure, and 1$\times 10^{-5}$ eV for the total energy. The structural optimization was performed under constant pressure using the Broydon-Fletcher-Goldfarb-Shanno (BFGS) algorithm ~\cite{broyden1970convergence1,broyden1970convergence2,goldfarb1970family,fletcher1970new,shanno1970optimal} with variable cell shape. The calculations of phonon spectra were carried out using the finite displacement approach as implemented in the PHONOPY code~\cite{togo2008first,togo2015first}.

\begin{figure}
\begin{center}
\includegraphics[width=3.5in]{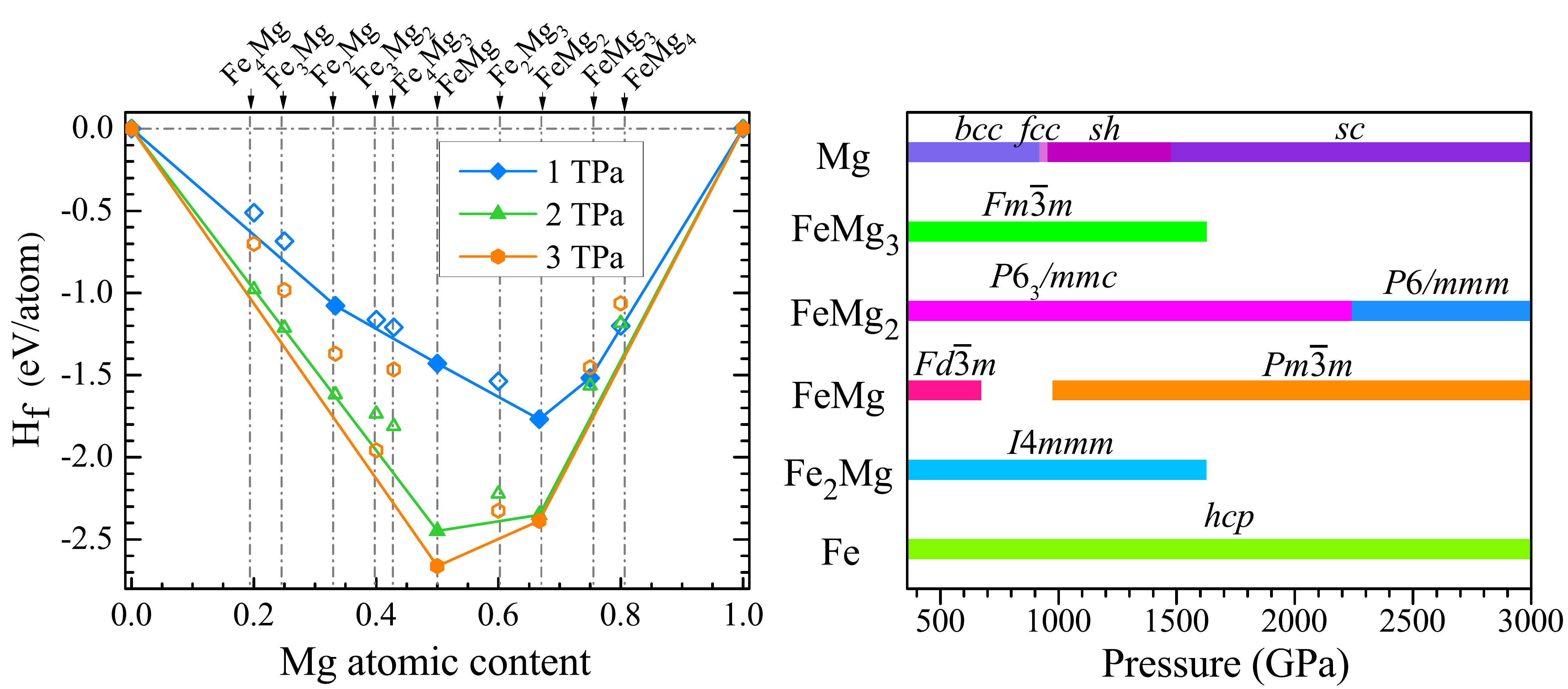}
\end{center}
\caption{Stability of Fe-Mg compounds. (a) Convex hull diagrams of the Fe-Mg compounds at exoplanetary pressures. (b) Pressure-composition phase diagram of the Fe-Mg system.}	
\end{figure}

\section{RESULTS AND DISCUSSION }

\subsection{Phase stability}
To obtain low-enthalpy structures in the Fe-Mg system, we performed an extensive search in the compositional space of Fe$_{x}$Mg$_{y}$ (x, y = 1$\sim$4) with maximum simulation cells containing up to 32 atoms at pressures of 1TPa, 2TPa, and 3TPa. The thermodynamic stability of Fe$_{x}$Mg$_{y}$ compounds was assessed by computing the formation enthalpies from the enthalpies of the elementary Fe and Mg in their stable phases at the same pressures. Specifically, the enthalpy of formation per atom (H$_f$) for a Fe$_{x}$Mg$_{y}$ phase is obtained as:
\begin{equation}
H_f = \frac{H_{Fe_{x}Mg_{y}}- (xH_{Fe}+yH_{Mg})}{x+y}
\end{equation}
Both elementary Fe and Mg exhibit multiple allotropes under pressure ~\cite{bassett1987mechanism,vattre2016polymorphism,zhu2013novel,li2010crystal}. Experimental and theoretical efforts have established well their phase diagrams. Here, the simple hexagonal (\emph{sh}) structured Mg and \emph{hcp}-Fe ground states are used as references at 1TPa. The simple cubic (\emph{sc}) Mg and \emph{hcp}-Fe are used as references at 2TPa and 3TPa. Fig. 1(a) depicts the Fe-Mg system\textquotesingle s convex hulls constructed using H$_f$. It is shown that four stoichiometric Fe$_{x}$Mg$_{y}$ phases, i.e., Fe$_2$Mg, FeMg, FeMg$_2$, and FeMg$_3$ are thermodynamically stable.
We construct the pressure-composition phase diagram in Fig. 1(b) from 360GPa, the upper limit for the pressure considered in Ref.\cite{gao2019iron}, to 3TPa. One observes that FeMg$_3$ and Fe$_2$Mg become unstable above 1590 GPa and 1625 GPa, respectively. FeMg$_2$ has two stable phases within the pressure range of our interest, with the phase transition occurring at 2241 GPa. At pressures below 675 GPa, FeMg has a stable phase with $Fd\bar{3}m$ symmetry, while at pressures above 976 GPa it stabilizes in a cubic lattice with $Pm\bar{3}m$ symmetry. All crystallographic parameters of the stable structures are listed in Supplementary Table S1 and Table S2.

\begin{figure}
\begin{center}
\includegraphics[width=3.5in]{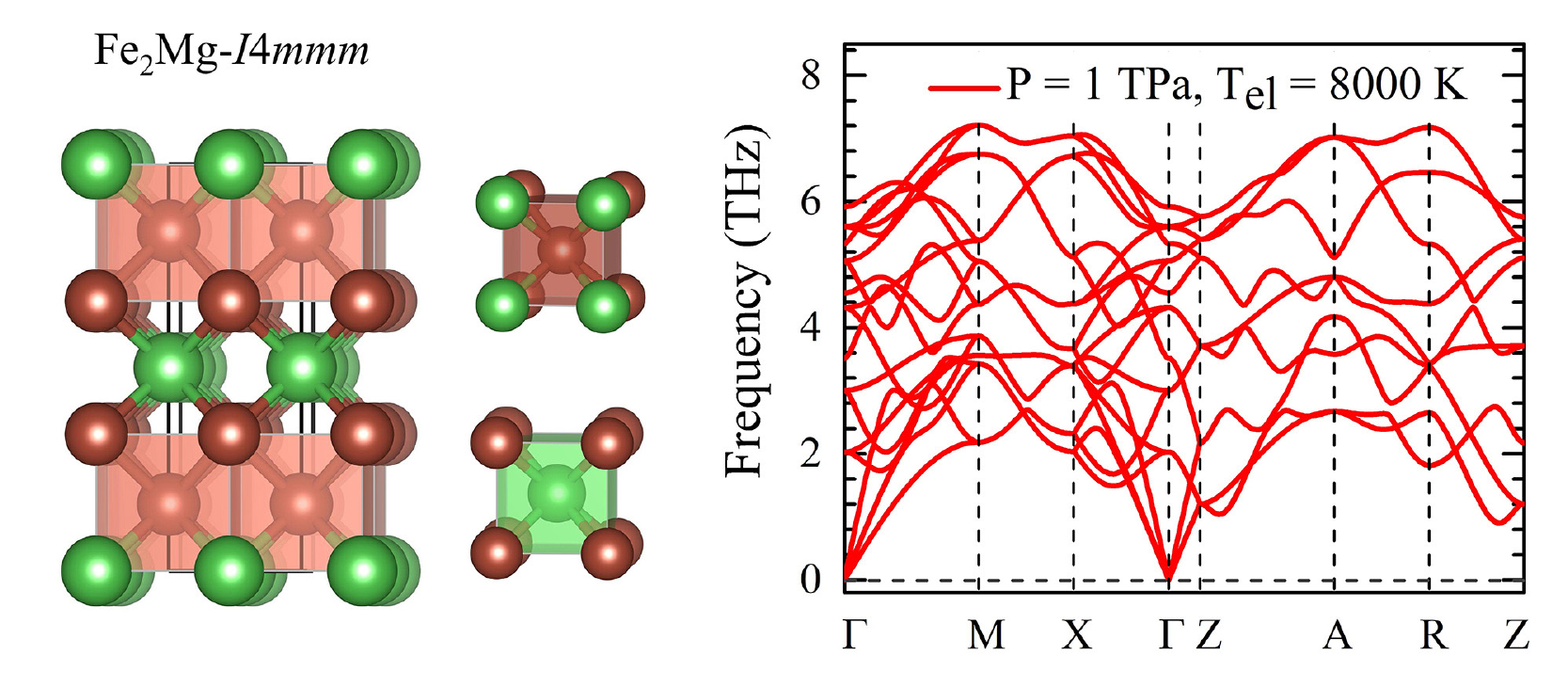}
\end{center}
\caption{Crystal structure and phonon dispersion of $I$4$/mmm$  Fe$_2$Mg. Fe and Mg atoms are indicated by brown and green balls, respectively.}	
\end{figure}
\subsection{Geometries and phonon stabilities}
\textbf{Fe$_2$Mg.} This Fe-rich phase forms a tetragonal structure with $I$4$/mmm$  symmetry (Fig. 2), which is the standard ground-state structure of binary compounds with A$_2$B stoichiometry at high pressures, e.g., Fe$_2$O~\cite{weerasinghe2015computational} and Al$_2$S~\cite{shao2020exotically}. In this structure, both Fe and Mg locate at the centers of the face-shared cube, but the difference is that each Fe is coordinated to 4 Fe and 4 Mg, while each Mg is bonded to 8 Fe. Interestingly, this structure was found to be stable from 220GPa to 360GPa by Gao \emph{et al}. Here we show that it can withstand high pressures up to 1625GPa. At higher pressures, it will decompose into FeMg and Fe. The phonon spectrum shown in Fig. 2 confirms that it is dynamically stable at 1TPa with an electron temperature (T$_{el}$) of 8000K. Generally, the temperature at the core-mantle boundary of a super-Earth falls within the range from 4000K to 10000K~\cite{van2019mass}. Therefore T$_{el}$ = 8000K is a reasonable choice. Nevertheless, the phonon spectra with T$_{el}$ = 0 K and T$_{el}$ = 3000 K are also presented in Fig. S1, showing no imaginary frequencies in the entire Brillouin zone.

\textbf{FeMg.} From 360GPa to 675GPa, the $Fd\bar{3}m$ phase previously identified in Ref.\cite{gao2019iron} is the ground state. The $Fd\bar{3}m$ phase has a BCC-like crystal structure such that each atom has 50\% of the nearest neighbour sites occupied by atoms of the same kind. From 976GPa to 3TPa, we find FeMg transform into the CsCl-type (B2) structure with $Pm\bar{3}m$ symmetry (see Fig. 3). In the pressure range from 675GPa to 976 GPa, FeMg decomposes to FeMg$_2$ and Fe$_2$Mg, which leaves a gap in the stability bar shown in Fig. 1. The dynamic stability of $Pm\bar{3}m$  FeMg at 1TPa, 2TPa, and 3TPa is verified by the absence of imaginary frequencies in the phonon dispersion, as shown in Fig. 3. Phonon dispersions with T$_{el}$= 0 K and T$_{el}$ = 3000 K are shown in Fig. S2.

\begin{figure}
\begin{center}
\includegraphics[width=3.5in]{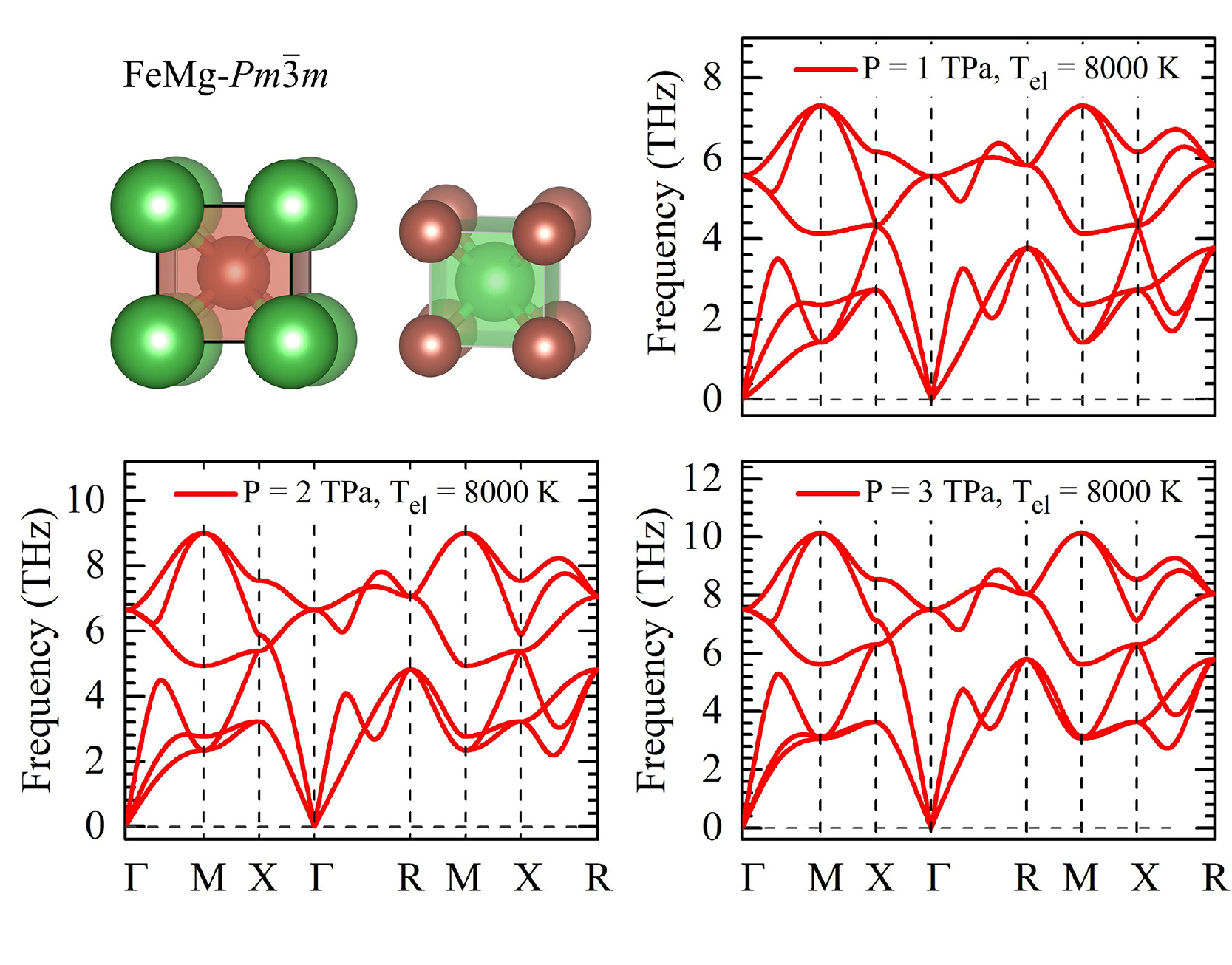}
\end{center}
\caption{Crystal structure and phonon spectra of $Pm\bar{3}m$ FeMg. Fe and Mg atoms are indicated by brown and green balls, respectively.}	
\end{figure}

\textbf{FeMg$_2$.} FeMg$_2$ adopts the hexagonal $P$6$_3$$/mmc$  structure at 1TPa and 2TPa. Each Fe in this phase is coordinated by five Mg, forming a Fe-centered face-sharing tetrahedron as shown in Fig. 4(a). While half of Mg forms the same polyhedra as Fe, Mg\textquotesingle s remaining half forms an isolated chain along the z-direction. At 2241 GPa, the $P$6$_3$$/mmc$ -FeMg$_2$ transforms into a hexagonal structure with $P$6$/mmm$  symmetry. $P$6$/mmm$-FeMg$_2$ features alternating triangular Fe layers with hexagonal Mg layers, as shown in Fig. 4(b). A similar structure was found in BaO$_2$, which was synthesized at $\sim$49.4GPa~\cite{efthimiopoulos2010structural}, despite the triangular layers formed by Ba and hexagonal layers formed by O being distorted. In this structure, each Fe is bonded to 12 Mg to form a hexagonal prism, while each Mg is coordinated to 3 Mg and 6 Fe to form polyhedra, as shown in Fig. 4(b). Phonon calculations show that the $P$6$_3$$/mmc$-FeMg$_2$ is dynamically stable at both 1TPa and 2TPa, see Fig. 4(a). The phonon dispersions with T$_{el}$ = 0 K and T$_{el}$ = 3000 K can be found in Fig. S3. At 3TPa, an imaginary frequency appears along the M-L path when the electronic temperature is 8000K (see Fig. 4(b)). However, it is dynamically stable at lower electronic temperatures (e. g., 3000K) (see Fig. S4).

\begin{figure}
\begin{center}
\includegraphics[width=3.5in]{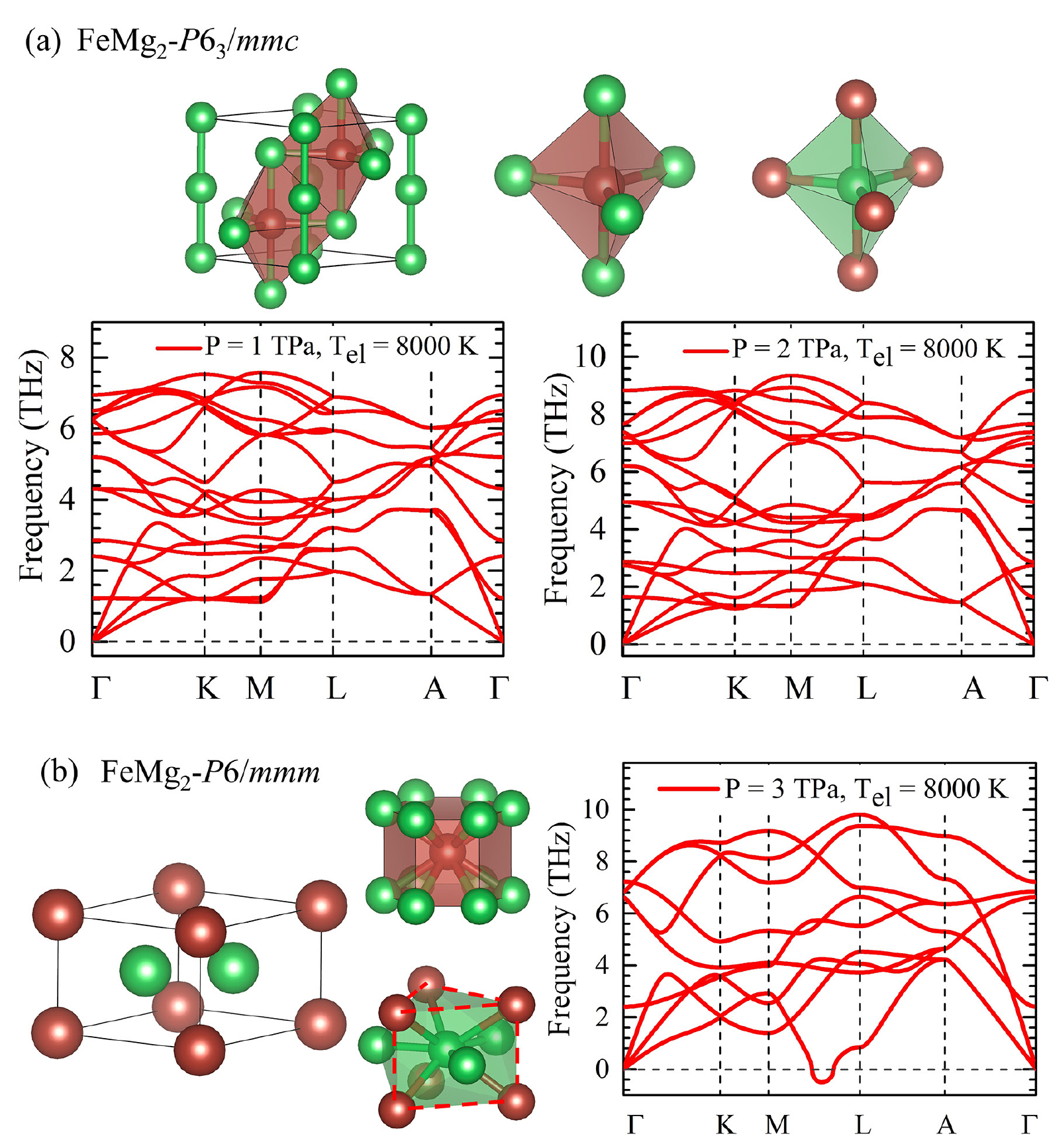}
\end{center}
\caption{(a) Crystal structure and phonon dispersions of $P$6$_3$$/mmc$  FeMg$_2$. (b) Crystal structure and phonon dispersion of $P$6$/mmm$  FeMg$_2$. Fe and Mg atoms are indicated by brown and green balls, respectively. }	
\end{figure}

\textbf{FeMg$_3$.} This phase exhibits a cubic structure with the $Fm\bar{3}m$ symmetry. It is composed of face-shared cubes with Fe/Mg being the central atoms, as shown in Fig. 5(a). It was reported that $Fm\bar{3}m$ FeMg$_3$ is stable within the pressure range from 307GPa to 360GPa. Our results reveal that this phase is stable below 1590GPa. At higher pressures, it will decompose into FeMg$_2$ and Fe. Phonon calculations show that it is dynamically unstable with low electron temperatures (see Fig. S4), while at electronic temperatures of 8000K, it becomes stable.

\begin{figure}
\begin{center}
\includegraphics[width=3.5in]{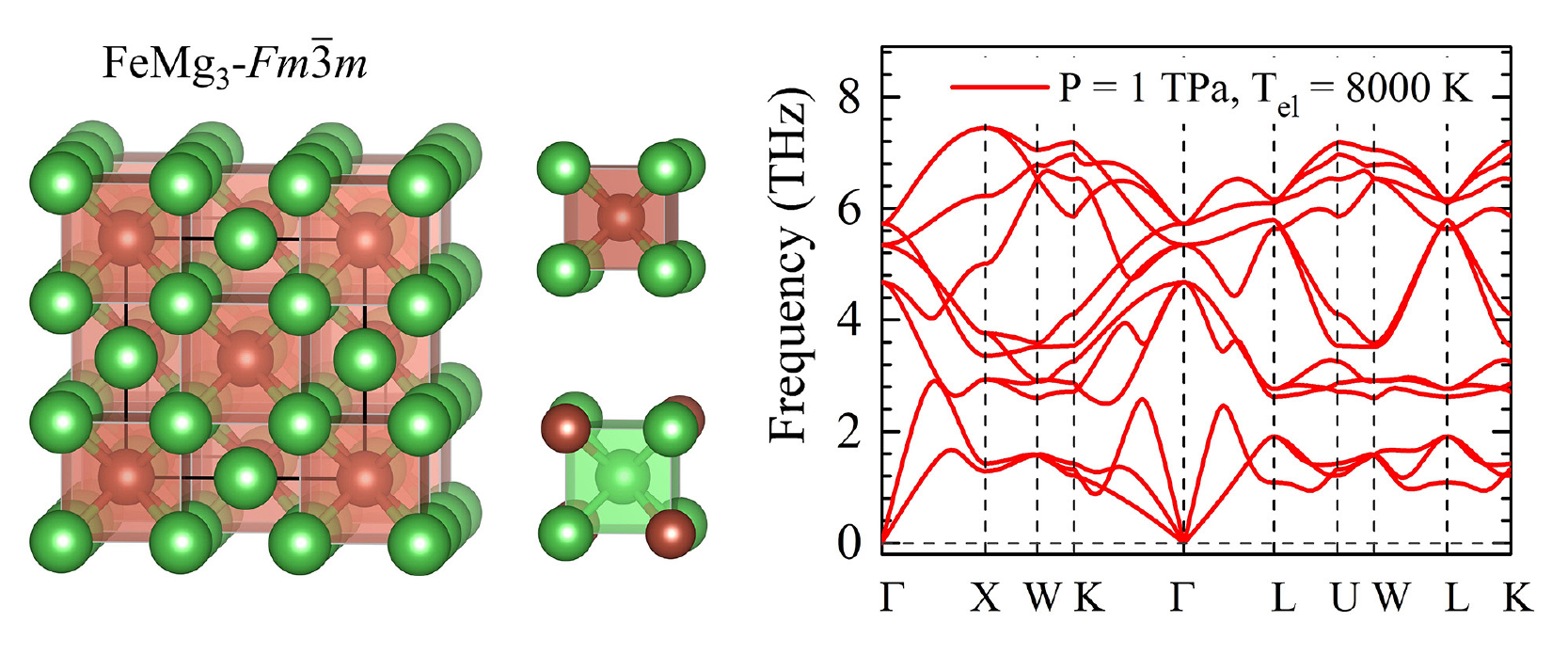}
\end{center}
\caption{Crystal structure and phonon spectrum of $Fm\bar{3}m$ FeMg$_3$. Fe and Mg atoms are indicated by brown and green balls, respectively.  }	
\end{figure}

\subsection{Local packing motifsm}
In addition to the stable structures, we also predict hundreds of metastable structures in the Fe-Mg system up to 3TPa. Since these are 0 K calculations, these low enthalpy metastable structures may become stable at elevated temperatures. In this respect, we also investigate the geometric features of those Fe$_{x}$Mg$_{y}$ phases with relative enthalpies (H$_d$) higher than the convex hull by 0.8 eV/atom ($\sim$9000K) to reveal the Fe-Mg system\textquotesingle s overall structural behavior at high pressures. The cluster alignment method~\cite{sun2016crystal}, which has successfully determined the crystal genes in crystals, glasses, and liquids, is adopted to identify these structures\textquotesingle ~packing motifs. We first align the Fe-centered clusters as extracted from the low-enthalpy Fe$_{x}$Mg$_{y}$ phases against six template motifs, as shown in the right panel of Fig. 6. The template motifs include FCC, BCC, HCP, OCT (octahedron), and BCT (body-centered tetragonal), which are the most popular motifs found in the Fe-O~\cite{weerasinghe2015computational} and Mg-O systems~\cite{zhu2013novel,niu2015prediction}. We can determine the structure\textquotesingle s building block in light of the alignment score, which describes the deviation of an as-extracted cluster from the perfect template. The alignment score criterion is set to be 0.125, allowing a small distortion of the crystal structures\textquotesingle  ~ideal motifs.

Figure 6 shows the relative enthalpies of the stable and metastable phases with respect to the convex hull as functions of their volumes. The local packing motifs are indicated with different symbols, and colors represent the Mg fraction. As shown in Fig. 6, when Fe and Mg atomic fractions are comparable, most Fe$_x$Mg$_y$ phases tend to adopt a single BCC motif. With high Fe or Mg content, different structural motifs can co-exist. At 360GPa, the averaged atomic volume increases with increasing Mg concentration. However, at 2TPa and 3TPa, the average atomic volumes decrease with increasing Mg concentration. At 1TPa, different Mg concentrations lead to similar averaged atomic volumes.

\begin{figure}
\begin{center}
\includegraphics[width=3.5in]{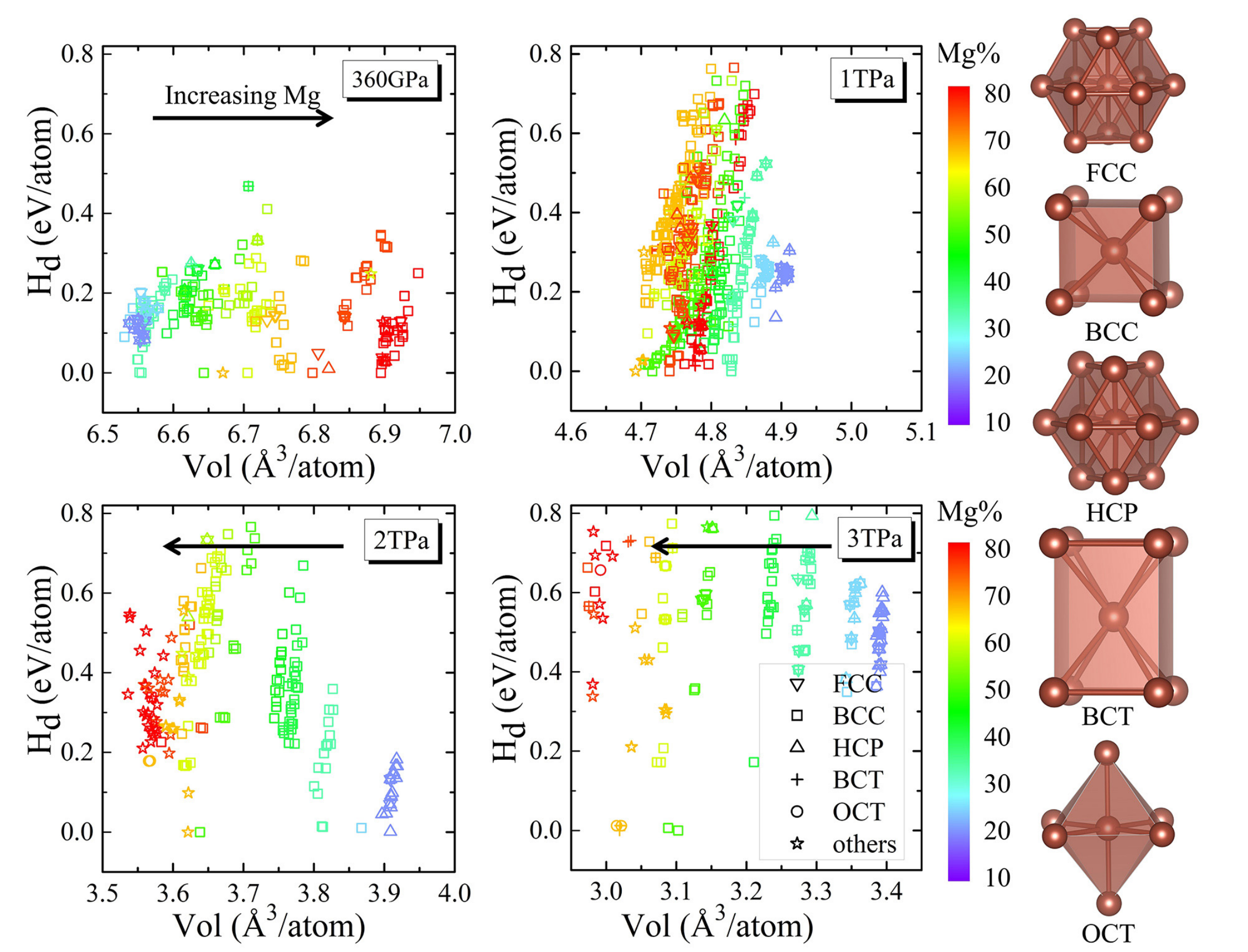}
\end{center}
\caption{The relative enthalpies of low-enthalpy Fe$_x$Mg$_y$ structures as a function of their volumes, where the symbols represent the local packing motifs, the colors denote Mg\textquotesingle s atomic content. The label \textquotesingle others\textquotesingle~  indicates a Fe-centered cluster with all six templates\textquotesingle ~lowest alignment scores higher than 0.125.  }	
\end{figure}

To understand the change of volume-composition relations, we investigate the compression behavior of elementary Fe and Mg phases under ultra-high pressures. We plot in Fig. 7 the pressure-volume relations for several Fe and Mg crystal structures. The solid lines are the fitting results of the third-order Birch-Murnaghan equation of state (EOS)~\cite{birch1978finite}. As shown in Fig. 7, all Fe allotropes have smaller atomic volume than Mg phases at pressures smaller than 0.6 TPa. In this range the atomic volume difference between elementary Fe and Mg decreases with the increasing pressure. Then, the volumes of two elements become similar from 0.6TPa to 0.9TPa. At pressures higher than 0.9TPa, the atomic volume of Fe allotropes becomes larger than those of Mg allotropes, and the volume difference increases with the increasing pressure. It is interesting to note that the atomic volume difference between Fe and Mg is more than one order of magnitude larger at ambient pressure than the one at ultra-high pressures (see Fig. 7 inset). Such a dramatical change of Fe/Mg volumes difference with respect to the pressure can explain the pressure-induced formation of Fe-Mg compounds. Under ambient pressure, the volume difference between Fe and Mg is so large that they are hardly miscible. With increasing pressure, Mg is more compressible than Fe, as evidenced by the volume difference reduction and volume crossover under pressure, leading to the formation of Fe-Mg compounds and different Fe-Mg motifs.

\begin{figure}[h]
\begin{center}
\includegraphics[width=3.4in]{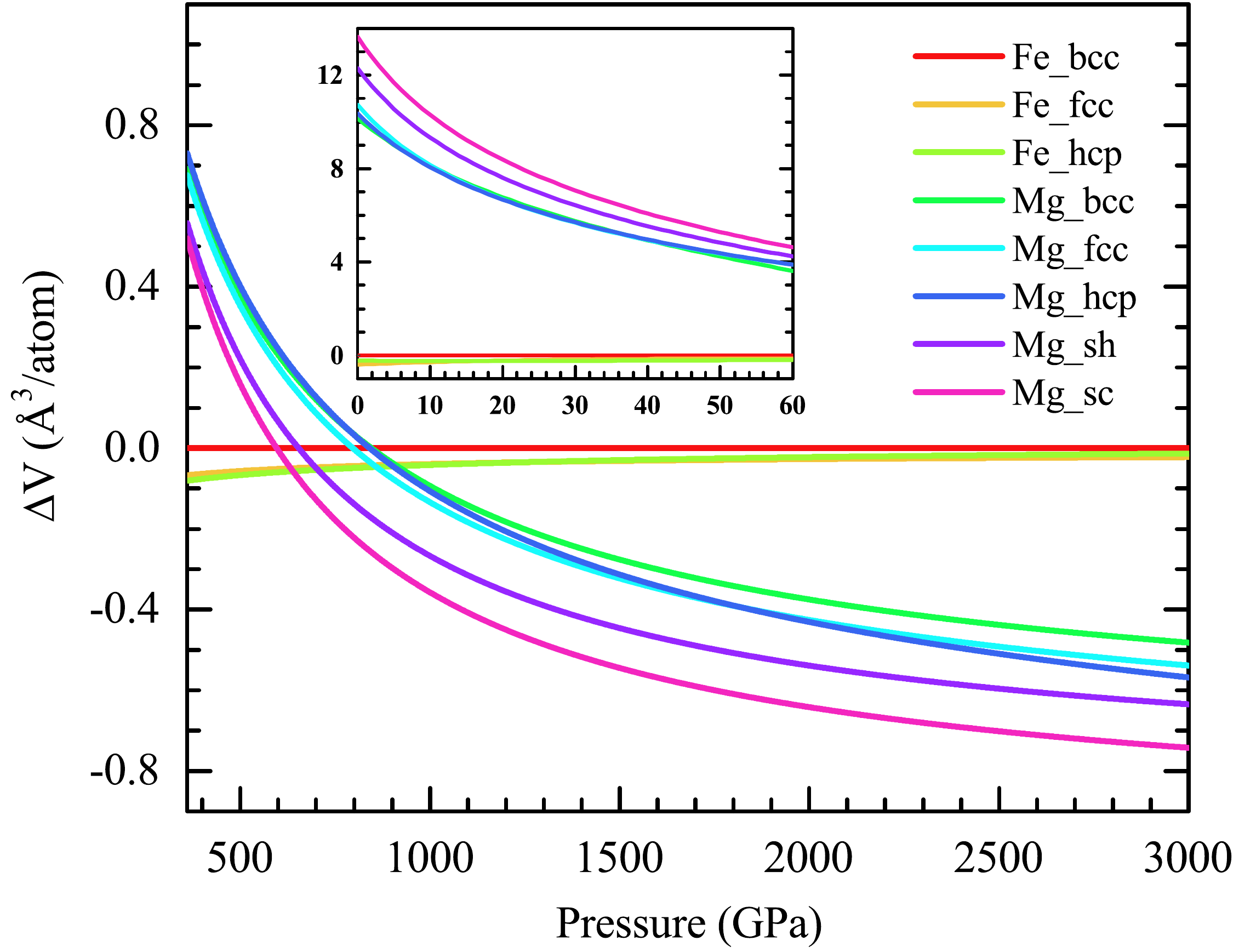}
\end{center}
\caption{Relative volume as a function of pressure for elementary Fe, Mg, and stable Fe$_x$Mg$_y$ phases. The inset shows the same at low pressures.  }	
\end{figure}
Our results suggest that stable stoichiometric Fe-Mg compounds should exist at extreme conditions of Super-Earths interiors, whether in the solid cores of those with few Earth masses (M$_{\oplus}$) or the mantle of heavier ones with more than 8M$_{\oplus}$[29]. From 400GPa and 1.6TPa, abundant stoichiometric compounds and Mg-Fe solid solutions should exist because Fe and Mg have similar atomic volumes, promoting their inter-mixing. Above 1.6TPa, the atomic volume difference is significant, again decreasing their inter-solubility. Only $\varepsilon$-Fe and BCC-like Fe$_2$Mg remain in Fe-rich stoichiometries, forming the basis for a euthetic sub-system in the cores of Super-Earths with few Earth masses. Such highly pressure-dependent solubility behavior may result in Super-Earth interiors with more complex layered structures than modeled so far~\cite{umemoto2017phase}.
~\\
~\\
~\\

\section{CONCLUSION}
In summary, we identified several stable stoichiometric phases in the Fe-Mg system under exoplanetary interior pressures using the efficient AGA search method combined with DFT calculations. In addition to the stable structures, we also predicted a significant number of metastable Fe$_x$Mg$_y$ structures with low enthalpies. The cluster alignment analysis reveals that all stable and metastable Fe-Mg compounds prefer a BCC packing motif at high pressures. Our study provides a more comprehensive structure database to support future investigations of the high-pressure behavior of Fe-Mg compounds.

\begin{acknowledgments}
Work at Xiamen University was supported by the National Natural Science Foundation of China (11874307). Work at Iowa State University and Columbia University was supported by the National Science Foundation awards EAR-1918134 and EAR-1918126. Work at the University of Science and Technology of China was supported by the National Natural Science Foundation of China (11574284 $\&$ 11774324) and the Supercomputing Center of USTC.
\end{acknowledgments}


\begin{thebibliography}{0}%
\makeatletter
\providecommand \@ifxundefined [1]{%
 \@ifx{#1\undefined}
}%
\providecommand \@ifnum [1]{%
 \ifnum #1\expandafter \@firstoftwo
 \else \expandafter \@secondoftwo
 \fi
}%
\providecommand \@ifx [1]{%
 \ifx #1\expandafter \@firstoftwo
 \else \expandafter \@secondoftwo
 \fi
}%
\providecommand \natexlab [1]{#1}%
\providecommand \enquote  [1]{``#1''}%
\providecommand \bibnamefont  [1]{#1}%
\providecommand \bibfnamefont [1]{#1}%
\providecommand \citenamefont [1]{#1}%
\providecommand \href@noop [0]{\@secondoftwo}%
\providecommand \href [0]{\begingroup \@sanitize@url \@href}%
\providecommand \@href[1]{\@@startlink{#1}\@@href}%
\providecommand \@@href[1]{\endgroup#1\@@endlink}%
\providecommand \@sanitize@url [0]{\catcode `\\12\catcode `\$12\catcode
  `\&12\catcode `\#12\catcode `\^12\catcode `\_12\catcode `\%12\relax}%
\providecommand \@@startlink[1]{}%
\providecommand \@@endlink[0]{}%
\providecommand \url  [0]{\begingroup\@sanitize@url \@url }%
\providecommand \@url [1]{\endgroup\@href {#1}{\urlprefix }}%
\providecommand \urlprefix  [0]{URL }%
\providecommand \Eprint [0]{\href }%
\providecommand \doibase [0]{https://doi.org/}%
\providecommand \selectlanguage [0]{\@gobble}%
\providecommand \bibinfo  [0]{\@secondoftwo}%
\providecommand \bibfield  [0]{\@secondoftwo}%
\providecommand \translation [1]{[#1]}%
\providecommand \BibitemOpen [0]{}%
\providecommand \bibitemStop [0]{}%
\providecommand \bibitemNoStop [0]{.\EOS\space}%
\providecommand \EOS [0]{\spacefactor3000\relax}%
\providecommand \BibitemShut  [1]{\csname bibitem#1\endcsname}%
\let\auto@bib@innerbib\@empty
\end{thebibliography}%


\begin{thebibliography}{10}

\bibitem{gao2019iron}
P.~Gao, C.~Su, S.~Shao, S.~Wang, P.~Liu, S.~Liu, and J.~Lv, New J. Chem.{ \bf
  43}, 17403--17407 (2019).

\bibitem{okamoto1990binary}
H.~Okamoto, T.~Massalski, et~al.
\newblock {\em Binary alloy phase diagrams}.
\newblock  (1990).

\bibitem{jaouen1989ion}
C.~Jaouen, J.~Delafond, N.~Junqua, and P.~Goudeau, Nucl. Instrum. Methods Phys.
  Res., Sect. B{ \bf 43}, 34--40 (1989).

\bibitem{yelsukov2005mechanism}
E.~Yelsukov, G.~Dorofeev, and A.~Ulyanov, Czech. J. Phys.{ \bf 55}, 913--921
  (2005).

\bibitem{dubrovinskaia2004iron}
N.~Dubrovinskaia, L.~Dubrovinsky, and C.~McCammon, J. Phys.: Condens. Matter{
  \bf 16}, S1143 (2004).

\bibitem{dubrovinskaia2005beating}
N.~Dubrovinskaia, L.~Dubrovinsky, I.~Kantor, W.~A. Crichton, V.~Dmitriev,
  V.~Prakapenka, G.~Shen, L.~Vitos, R.~Ahuja, B.~Johansson, et~al., Phys. Rev.
  Lett.{ \bf 95}, 245502 (2005).

\bibitem{kadas2009stability}
K.~K{\'a}das, L.~Vitos, B.~Johansson, and R.~Ahuja, Proc. Natl. Acad. Sci.
  U.S.A{ \bf 106}, 15560--15562 (2009).

\bibitem{li2018mg}
Y.~Li, L.~Vo{\v{c}}adlo, D.~Alf{\`e}, and J.~Brodholt, Phys. Earth Planet.
  Inter.{ \bf 274}, 218--221 (2018).

\bibitem{deaven1995molecular}
D.~M. Deaven and K.-M. Ho, Phys. Rev. Lett.{ \bf 75}, 288 (1995).

\bibitem{foiles1986embedded}
S.~Foiles, M.~Baskes, and M.~S. Daw, Phys. Rev. B{ \bf 33}, 7983 (1986).

\bibitem{banerjea1988origins}
A.~Banerjea and J.~R. Smith, Phys. Rev. B{ \bf 37}, 6632 (1988).

\bibitem{brommer2006effective}
P.~Brommer and F.~G{\"a}hler, Philos. Mag.{ \bf 86}, 753--758 (2006).

\bibitem{brommer2007potfit}
P.~Brommer and F.~G{\"a}hler, Modell. Simul. Mater. Sci. Eng.{ \bf 15}, 295
  (2007).

\bibitem{giannozzi2009quantum}
P.~Giannozzi, S.~Baroni, N.~Bonini, M.~Calandra, R.~Car, C.~Cavazzoni,
  D.~Ceresoli, G.~L. Chiarotti, M.~Cococcioni, I.~Dabo, et~al., J. Phys.:
  Condens. Matter{ \bf 21}, 395502 (2009).

\bibitem{giannozzi2017advanced}
P.~Giannozzi, O.~Andreussi, T.~Brumme, O.~Bunau, M.~B. Nardelli, M.~Calandra,
  R.~Car, C.~Cavazzoni, D.~Ceresoli, M.~Cococcioni, et~al., J. Phys.: Condens.
  Matter{ \bf 29}, 465901 (2017).

\bibitem{broyden1970convergence1}
C.~G. Broyden, IMA J. Appl. Math.{ \bf 6}, 76--90 (1970).

\bibitem{broyden1970convergence2}
C.~G. Broyden, IMA J. Appl. Math.{ \bf 6}, 222--231 (1970).

\bibitem{goldfarb1970family}
D.~Goldfarb, Math. Comput.{ \bf 24}, 23--26 (1970).

\bibitem{fletcher1970new}
R.~Fletcher, Comput. J.{ \bf 13}, 317--322 (1970).

\bibitem{shanno1970optimal}
D.~F. Shanno and P.~C. Kettler, Math. Comp.{ \bf 24}, 647--656 (1970).

\bibitem{togo2008first}
A.~Togo, F.~Oba, and I.~Tanaka, Phys. Rev. B{ \bf 78}, 134106 (2008).

\bibitem{togo2015first}
A.~Togo and I.~Tanaka, Scr. Mater.{ \bf 108}, 1--5 (2015).

\bibitem{bassett1987mechanism}
W.~Bassett and E.~Huang, Science{ \bf 238}, 780--783 (1987).

\bibitem{vattre2016polymorphism}
A.~Vattr{\'e} and C.~Denoual, J. Mech. Phys. Solids{ \bf 92}, 1--27 (2016).

\bibitem{zhu2013novel}
Q.~Zhu, A.~R. Oganov, and A.~O. Lyakhov, Phys. Chem. Chem. Phys.{ \bf 15},
  7696--7700 (2013).

\bibitem{li2010crystal}
P.~Li, G.~Gao, Y.~Wang, and Y.~Ma, J. Phy. Chem. C{ \bf 114}, 21745--21749
  (2010).

\bibitem{weerasinghe2015computational}
G.~L. Weerasinghe, C.~J. Pickard, and R.~Needs, J. Phys.: Condens. Matter{ \bf
  27}, 455501 (2015).

\bibitem{shao2020exotically}
S.~Shao, W.~Zhu, J.~Lv, Y.~Wang, Y.~Chen, and Y.~Ma, npj Comput. Mater.{ \bf
  6}, 11 (2020).

\bibitem{van2019mass}
A.~P. Van Den~Berg, D.~A. Yuen, K.~Umemoto, M.~H. Jacobs, and R.~Wentzcovitch,
  Icarus{ \bf 317}, 412--426 (2019).

\bibitem{efthimiopoulos2010structural}
I.~Efthimiopoulos, K.~Kunc, S.~Karmakar, K.~Syassen, M.~Hanfland, and
  G.~Vajenine, Phys. Rev. B{ \bf 82}, 134125 (2010).

\bibitem{sun2016crystal}
Y.~Sun, F.~Zhang, Z.~Ye, Y.~Zhang, X.~Fang, Z.~Ding, C.-Z. Wang, M.~I.
  Mendelev, R.~T. Ott, M.~J. Kramer, et~al., Sci. Rep.{ \bf 6}, 23734 (2016).

\bibitem{niu2015prediction}
H.~Niu, A.~R. Oganov, X.-Q. Chen, and D.~Li, Sci. Rep.{ \bf 5}, 18347 (2015).

\bibitem{birch1978finite}
F.~Birch, J. Geophys. Res. Solid Earth{ \bf 83}, 1257--1268 (1978).

\bibitem{umemoto2017phase}
K.~Umemoto, R.~M. Wentzcovitch, S.~Wu, M.~Ji, C.-Z. Wang, and K.-M. Ho, Earth
  Planet. Sci. Lett.{ \bf 478}, 40--45 (2017).



\end{thebibliography}

\end{document}